\documentstyle[aps,prbbib,twocolumn,epsf]{revtex}

\begin{document}
\draft
\title{Resonant Andreev reflections in superconductor-carbon-nanotube devices}

\author{Yadong Wei$^1$, Jian Wang$^1$, Hong Guo$^2$, Hatem Mehrez$^2$, 
and Christopher Roland$^3$}

\address{1. Department of Physics, The University of Hong Kong, 
Pokfulam Road, Hong Kong, China.\\
2. Center for the Physics of Materials and Department 
of Physics, McGill University, Montreal, PQ, Canada H3A 2T8.\\
3. Department of Physics, The North Carolina State University,
Raleigh, NC USA 27695.
}
\maketitle

\begin{abstract}
Resonant Andreev reflection through superconductor-carbon-nanotube devices 
was investigated theoretically with a focus on the superconducting proximity 
effect. Consistent with a recent experiment, we find that for high 
transparency devices on-resonance, the Andreev current is characterized 
by a large value and a resistance dip; low-transparency off-resonance devices 
give the opposite result. We also give evidence that the observed 
low-temperature transport anomaly may be a natural result of Andreev 
reflection process. 
\end{abstract}


The field of carbon nanotube research has recently entered a new phase with 
the fabrication of hybrid device structures, in which carbon nanotubes are 
contacted electrically with other materials\cite{tsukagoshi,zhang,yu,dai}. 
This is a crucial step, as a carbon nanotube-based microelectronics is only 
possible when finite nanotubes can be efficiently fabricated and coupled 
to external leads. 
%
%
In this Letter we report a theoretical analysis of a hybrid 
superconductor-nanotube junction which has been the subject of a recent 
experimental study\cite{dai}. The experimental device consists of a single-wall
metallic carbon nanotube (SWNT) bridging two superconducting electrodes.
By tuning the transparency of the device\cite{dai}, clear signals of Andreev
reflections\cite{andreev} were detected via changes in the sub-gap resistance
at a temperature of $T=4.2$K, while other transport anomalies were observed at
lower $T$. To date, there have been many theoretical and experimental 
studies of normal metal (N)/superconductor (S) interfaces on a mesoscopic 
scale\cite{beena1,lambert1,beena,ando,yeyati1,sun,wendin,ruitenbeek,kleinsasser,kastalsky,scheer}. 
However, no such analysis exists for molecular devices. By combining standard 
nonequilibrium Green's function techniques (NEGF)\cite{ando,yeyati1,sun} with 
a tight-binding model (TB) for the SWNT, we have analyzed quantum transport 
properties of SWNT-S junctions. Our results are consistent with the 
experimental data\cite{dai}.

Although the experimental device consisted of {\it two} SWNT-S junctions, the
data indicates that each of these junctions acts independently\cite{dai}.
Hence, we focus here on the somewhat simpler problem of a N-SWNT-S system,
leaving an analysis of the multiple Andreev reflections of a S-SWNT-S system 
for the future. Our theory proceeds by combining the NEGF with a standard 
TB model for the SWNT\cite{chico} such that the coupling of 
the nanotube to the N (left) and S (right) leads are included via their 
appropriate self-energies. By iterating the equation of motion\cite{jauho}, 
standard but tedious algebra\cite{sun} shows that there are two contributions 
to the electric current flowing through the device: {\it i.e.}, $I=I_A+I_1$. 
The Andreev current $I_A$, or sub-gap contribution, is given by $(\hbar=e=1)$:
\begin{equation}
I_A=\frac{1}{\pi}\int dE \left[ f_L(E-V)-f_L(E+V)\right] T_A(E),
\label{IA}
\end{equation}
where $f_{L,R}$ denote the Fermi functions of the left and right leads 
respectively, $E$ the electron energy, and $V$ the bias potential. The 
Andreev reflection probability $T_A(E)$ is given by:
\begin{equation}
T_A(E)\equiv \ Tr\left[ {\bf \Gamma}_L
{\bf G}_{12}(E){\bf \Gamma}_L {\bf G}_{12}^{\dagger}(E)\right],
\label{ta}
\end{equation}
where $\Gamma_{L,R}$ are the appropriate line-width functions describing 
the coupling of the SWNT to the respective leads. Here ${\bf G_{11}}$ and 
${\bf G_{12}}$ are the retarded Green's functions\cite{yeyati1,sun} of the 
SWNT which include the proper self energies of the leads. These are evaluated 
by direct matrix inversion\cite{hatem}. The remaining contribution 
to the current is given by:
\begin{eqnarray}
I_1 &=& \frac{1}{\pi}\int dE \ \ Tr
\left[ T_1(E) + T_2(E) + T_3(E)\right]
\nonumber \\
& & \times \left[f_L(E-V)-f_R(E)\right]\rho_R(E)
\label{i1}
\end{eqnarray}
where the density of states of the S lead is 
$\rho_R(E)=|E|/\sqrt{E^2-\Delta^2}$ for $|E|>\Delta$, and zero otherwise; 
and $\Delta$ is the gap energy of the superconductor. Here $T_1,\ T_2,\ T_3$ 
are transmission probabilities for different physical processes that are 
only nonzero when $|E|\ge \Delta$. Hence, these processes describe excitations
of the system. In particular, $T_1={\bf \Gamma}_L{\bf G}_{11}{\bf \Gamma}_R 
{\bf G}_{11}^{\dagger}$ gives the familiar tunneling current; 
$T_2={\bf \Gamma}_L{\bf G}_{12} {\bf \Gamma}_R {\bf G}_{12}^{\dagger}$ 
denotes the branch crossing process of Blonder-Tinkham-Klapwijk 
theory\cite{bjk}; and 
$T_3=-(\Delta/|E|)[{\bf \Gamma}_L{\bf G}_{11} {\bf \Gamma}_R 
{\bf G}_{12}^{\dagger}+H.C.]$ describes the Cooper pair formation 
(or annihilation) inside the superconducting lead by an incoming electron
(or hole). Clearly, at zero temperature when bias voltage $|V|>\Delta$, all 
the processes will contribute to current; while for $|V|<\Delta$ only the 
Andreev current $I_A$ is nonzero. 

In our numerical calculations, we fixed the gap energy at $\Delta=1.45$ meV 
corresponding to that of Nb leads; the SWNT TB Hamiltonian is taken to be a 
nearest-neighbor $\pi -$orbital model with bond potential $V_{pp\pi }=-2.75$ 
eV, which is known to give a reasonable description of the electronic and 
transport properties of carbon nanotubes\cite{chico,marco}. We focus metallic
nanotubes of finite length $L$. The proximity of the nanotube to the 
superconductor is modeled through the coupling parameters $\Gamma_{L,R}$, 
which are treated as input parameters\cite{wan1}.

The solid lines of Fig. (1) show Andreev reflection coefficient $T_A(E)$ 
for a metallic (5,5) SWNT device of length\cite{foot1} $L=3\times 6+1=19$, and
for a (9,0) zigzag SWNT device of $L=49$, with coupling parameter 
$\Gamma_L=\Gamma_R=5\times 10^{-3}$a.u. (1 a.u. energy is 13.6eV). 
It is clear that resonant transmission with $T_A(E)=2.0$, {\it i.e., high 
device transparency}, dominates the transport at $E=0$. This may be understood 
as follows. It is well-known, that infinitely-long armchair nanotubes have two 
states crossing the Fermi level at $E_F=0$. On the other hand a finite-length 
isolated armchair nanotube has a gap between the two eigenstates near $E_F$, 
and this gap is minimized for tube length $L=3n+1$ where $n$ an 
integer\cite{rochefort1,hatem}. Therefore, when coupled to the device 
electrodes, which also adds a finite width to the levels, these nanotubes 
have two scattering states at $E_F$ giving rise to large transmission with 
$T_A=2$ as shown in Fig.1. The large $T_A$ for the zigzag tube is
also due to a resonance transmission through the nanotube states at the
Fermi level. The two other curves in Fig.1  show results for {\it asymmetric}
couplings and will be discussed later. Thus, for a N-SWNT-S system, the device 
transparency is critically determined by quantum resonance phenomenon which is
qualitatively different from the case of an infinite SWNT. As expected, 
high transparency devices have a larger $T_A$, since more electrons arriving 
at the SWNT-S interface increase the magnitude of the Andreev reflection 
process.
 
Figs.2 and 3 show the current-voltage characteristics, and the differential 
resistance $dV/dI$ for high- and low-transparency (5,5) SWNT devices, 
respectively. The I-V curves of Fig.2 are qualitatively consistent with 
and quantitatively close to the experimental data\cite{dai}. It is clear 
that a higher slope is observed for the I-V curves within the sub-gap range 
for on-resonance devices (dashed curve). This gives rise to an asymmetric 
resistance dip near bias voltage $V=0$, (see Fig.3a) with the asymmetry due 
to the finite gate voltage $V_g$. This resistance dip is simply a reflection 
of the high value of $T_A$ for on-resonance devices (solid line - Fig.1). The 
resistance dip has a value close to $h/(2\times 4e^2)=3.2K\Omega$ which is 
precisely the expected value of Andreev reflection processes in a SWNT-S 
junction with two transmitting modes. The experimental data\cite{dai} for 
two SWNT-S junctions connected in series actually gives a value close to 
$5.7K\Omega$, which is not far from the expected value of $6.4K\Omega$. 
This difference is perhaps due to parallel connection of three SWNTs bridging 
the superconductor electrodes in the experimental setup\cite{dai}.
When the device transparency is low, the differential resistance displays 
a large peak at $V=0$, as shown in Fig.3b. This is consistent both with 
the I-V curve of the low-transparency device shown in Fig. 2 and the 
experimental data\cite{dai}.

So far, the data presented have been for temperatures of $T=4.2$K, 
so that features reflecting smaller energy scales are completely washed 
out. However, at a lower temperature of $T=2$K, the experimental 
data\cite{dai} shows that a narrow peak emerges in the $dV/dI$ curves for 
zero bias, which is superimposed on the Andreev dip. Such anomalous behavior 
has previously been ascribed to the strong electron-electron interactions 
characteristic of Luttinger liquids. Surprisingly, our analysis shows that 
these features emerge naturally at lower temperatures, even within the 
context of a single-electron theory as presented here. This is shown in 
Fig.4, which illustrates the emergence of a narrow peak out of the overall 
Andreev dip as the temperature is lowered.

We can trace this low temperature anomaly to the basic physical process which
gives rise to the Andreev current. To demonstrate this, we neglect 
complications due to the molecular structure of the nanotubes and
assume that the resonant Andreev process is mediated by a single state at 
energy $E_o <\Delta$, {\it i.e.} we ``shrink'' the nanotube to a simple 
quantum well with a single level. This is qualitatively reasonable because 
the sub-gap energy scale is set by gap energy $\Delta$ which is much 
smaller than the level spacing of the nanotube we study\cite{rubio}.
Hence, we expect that only the two degenerate levels at the Fermi energy 
will contribute appreciably to the Andreev current. For this single-level 
case, the Green's functions are drastically simplified\cite{sun} so that 
equivalently, familiar scattering matrix theory can be applied\cite{beena1}. 
Near the resonance, the transmission amplitude in the 
normal state assumes the Breit-Wigner form, 
$t(E)=i\sqrt{\Gamma_L\Gamma_R}/(E-E_o+i\Gamma/2)$; and the reflection amplitude 
becomes $r(E)=1-i\Gamma_R/(E-E_o+i\Gamma/2)$; where 
$\Gamma\equiv \Gamma_L+\Gamma_R$ is the total line width. For these
amplitudes, Eq. (\ref{ta}) takes on the Breit-Wigner form:
\begin{equation}
T_A(E)=\frac{\Gamma_L^2\Gamma_R^2}{4\left[ 
E^2-E_o^2+\frac{\Gamma\delta\Gamma}{4} \right]^2 + \Gamma_L^2\Gamma_R^2
+E_o^2\left[ \Gamma^2+\delta\Gamma^2 \right] },
\label{Ta1}
\end{equation}
where $\delta\Gamma\equiv \Gamma_L-\Gamma_R$. This result allows us to draw 
several conclusions. For simplicity, we set the level at $E_o=0$, to obtain
\begin{equation}
T_A(E)=\frac{\Gamma_L^2\Gamma_R^2}{4\left(
E^2+\frac{\Gamma\delta\Gamma}{4}\right)^2+\Gamma_L^2\Gamma_R^2 }
\ .
\label{Ta11}
\end{equation}
These expressions reduce to the result of scattering matrix theory for 
N-Dot-S systems\cite{beena1} when we set $E=E_o=0$.  Eq.(\ref{Ta11})
indicates that if $\Gamma_L=\Gamma_R$ so that $\delta\Gamma=0$, then
resonant Andreev reflection occurs at $E=0$ with $T_A(E=0)=1$. On the other 
hand, if $\Gamma_L >\Gamma_R$ so that $\delta \Gamma > 0$, $T_A$ takes on 
a maximum value at $E=0$ but this maximum value is less than $1$. 
For nanotubes this situation is shown by the dotted lines of Fig.1. 
Furthermore, if $\Gamma_L < \Gamma_R$ such that $\delta\Gamma <0$, $T_A$ is 
characterized by two resonant peaks with $T_A=1$ at energies 
$E_{\pm}=\pm \sqrt{-\Gamma\delta\Gamma}/2$. For nanotubes this behavior is 
shown by the dashed lines of Fig.1. Hence, due to a split between the electron
and hole levels when the nanotube is in contact with a superconducting
lead, the Andreev coefficient in the sub-gap region can display different
behaviors. When $T_A(E)$ displays a double peak, the Andreev current $I_A$
shows a resistance anomaly such that a small peak develops inside the overall 
dip at low temperatures. At higher temperatures such as $4.2K$, the anomaly 
is smeared out, and  hence, not observable. One can also confirm that, 
qualitatively, the above conclusion holds for cases of nonzero gate voltage.

The analysis presented so far has been for (5,5) nanotubes with a length 
of $L=19$ unit cells, and (9,0) zigzag nanotubes\cite{foot1} with $L=49$. 
However, we expect our results to be general for other metallic nanotubes 
and lengths. Previous investigations showed\cite{hatem} 
that transport through {\it finite} armchair nanotubes differ qualitatively 
from the {\it infinite} length limit. In particular, if $L=3n+1$ where $n$ 
is an integer, the armchair tube has large conductance due to the crossing 
of scattering state energy levels at the Fermi energy. Other tube lengths 
produce much smaller conductance due to a gap between the scattering states. 
The inset of Fig.2 shows the I-V curve of N-SWNT-S devices with $L=17$
and $L=18$: these devices have very small currents because their device
transparency are drastically diminished by the energy gap between the 
scattering states of the SWNT. Experimentally, such an energy gap along 
with conductance oscillations on finite tubes have already been detected 
with scanning probes\cite{venema}.

In summary, we have investigated the sub-gap transport properties of
N-SWNT-S systems and our results are consistent with a number of
experimental observations. Specifically, the dependence of the Andreev current 
on the device transparency, the behavior of the differential resistance in 
the sub-gap region, as well as the observed low-temperature resistance 
anomaly may all be explained in terms of the proximity of the nanotubes with 
the superconducting lead. There are, however, still several issues which 
cannot be studied within the context of our model. First, the 
experimental data\cite{dai} showed a sensitive gate voltage dependence at 
very low temperatures on the order of $40$ mK, where the differential 
resistance anomaly could be a peak or a dip depending on the value of the 
gate voltage. This behavior has been considered as likely due to 
electron-electron interactions\cite{dai}. Second, there are experimental 
and theoretical evidence that nanotubes can have non-Fermi liquid 
behavior\cite{luttinger,bockrath}. It will be of great interest to 
investigate the 
situation of a non-Fermi liquid model of nanotubes in contact with a 
superconductor to see if other, finer features emerge. Another important 
problem is the detailed atomic structural analysis of the 
nanotube-superconductor interface which, to a large extent, controls the 
interface transparency. Finally, although we do not expect charge transfer 
to play a critical role in understanding the Andreev current for SWNT-S 
interfaces because of the superconducting gap, a more complete investigation 
of this delicate effect will certainly enhance our understanding of quantum 
transport for nanoscale devices.

{\bf Acknowledgments:} We gratefully acknowledge Prof. T.H. Lin and Dr.
Q.F. Sun for helpful discussions on the NEGF theory. We gratefully acknowledge 
financial support from NSERC of Canada and FCAR of Quebec (H.G); RGC grant 
(HKU 7215/99P) from the Hong Kong SAR (J.W.); ONR N00014-98-1-0597 and 
NASA NAG8-1479 (C.R.). We also thank the North Carolina Supercomputing Center
(NCSC) for significant amounts of computer time.


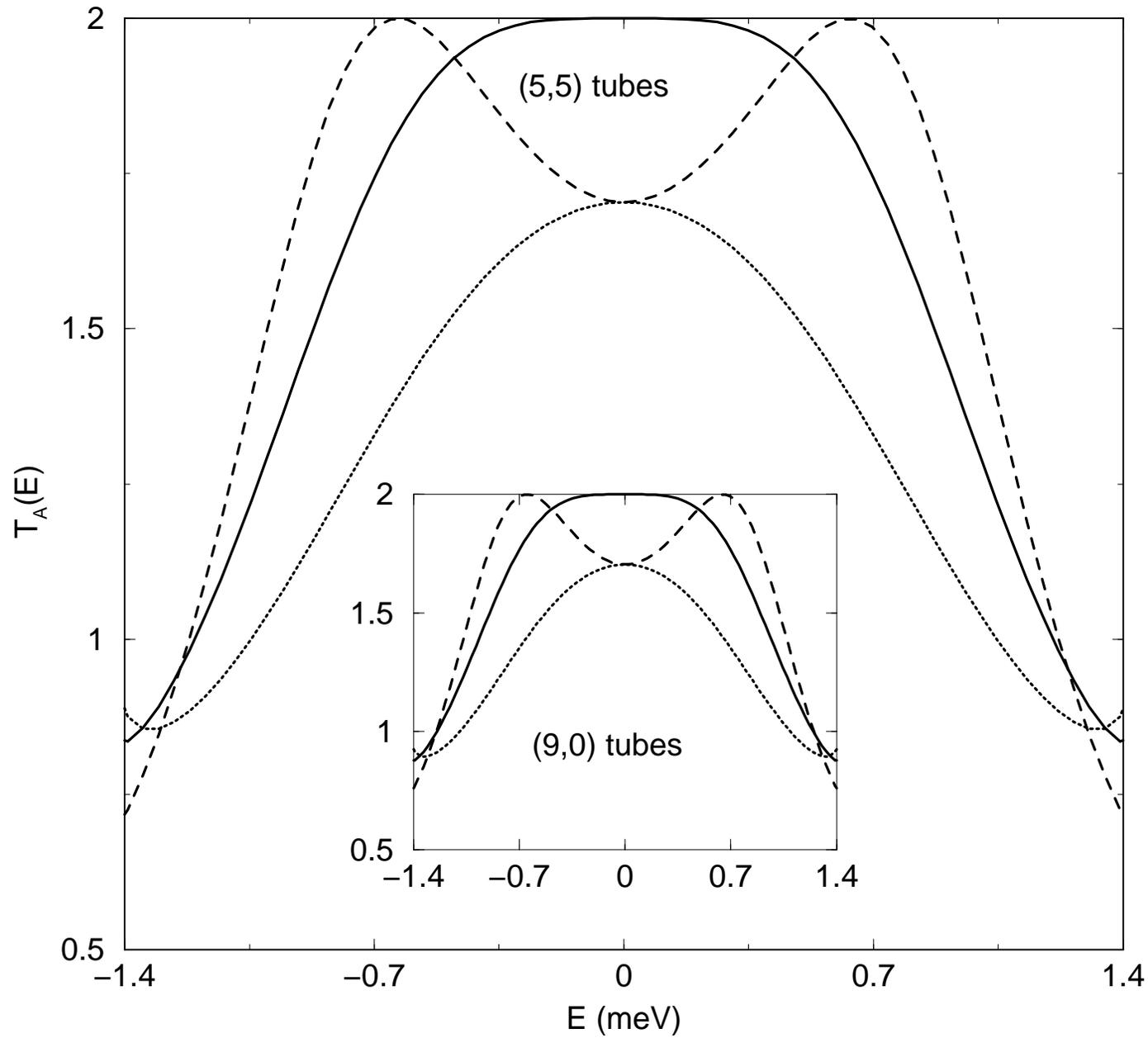
\begin{figure}
\caption{
Andreev reflection coefficient $T_A$ as a function of electron energy $E$.
Main graph: for devices consisting of (5,5) SWNT with length L=19 unit cells.
Inset: For (9,0) zigzag nanotube systems.  Other parameters are fixed as 
zero bias and zero gate voltages. Solid curve: 
$\Gamma_L=\Gamma_R=5\times 10^{-3}$; dotted curve: 
$\Gamma_L=6\times 10^{-3}$, $\Gamma_R=4\times 10^{-3}$; dashed curve: 
$\Gamma_L=4\times 10^{-3}$, $\Gamma_R=6\times 10^{-3}$. Here the
$\Gamma$s are measured in a.u..
}
\end{figure}

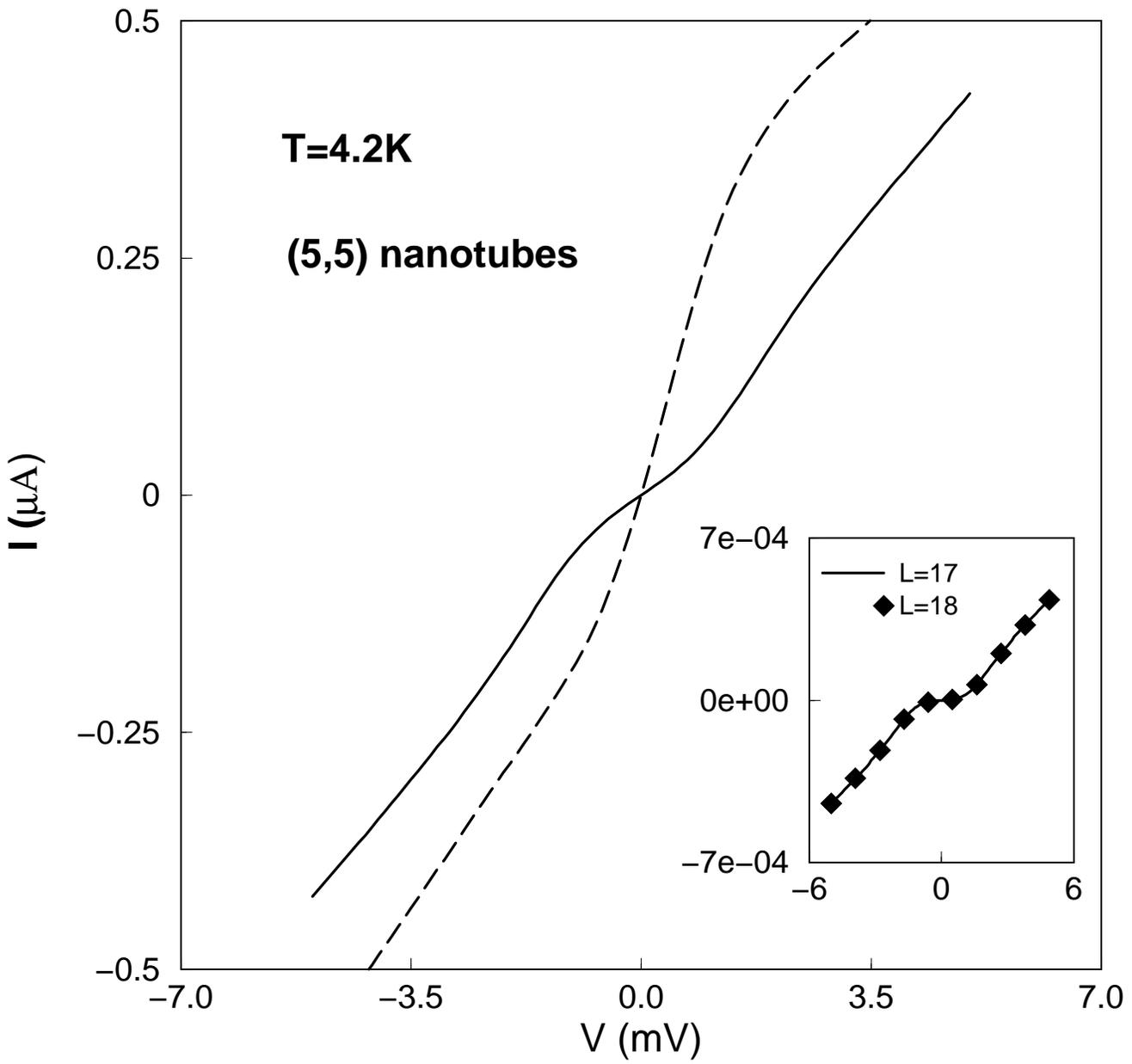
\begin{figure}
\caption{I-V curves for the N-SWNT-S device at temperature $4.2$K. 
The SWNT is a (5-5) metallic tube with length $L=19$ unit cells.
Solid curve is for a low transparency device (off resonance transmission) 
with parameters $\Gamma_L=\Gamma_R=0.8$ and $V_g=0$. Dashed curve is
for a high transparency device (on resonance transmission) with
$\Gamma_L=\Gamma_R=5\times 10^{-3}$ and $V_g=0.6$mV.
Inset: I-V curves (same units as the main graph)
for tubes with $L=17$ and $18$ unit cells with
$\Gamma_L=\Gamma_R=5\times 10^{-3}$ and $V_g=0.6$mV.
}
\end{figure}

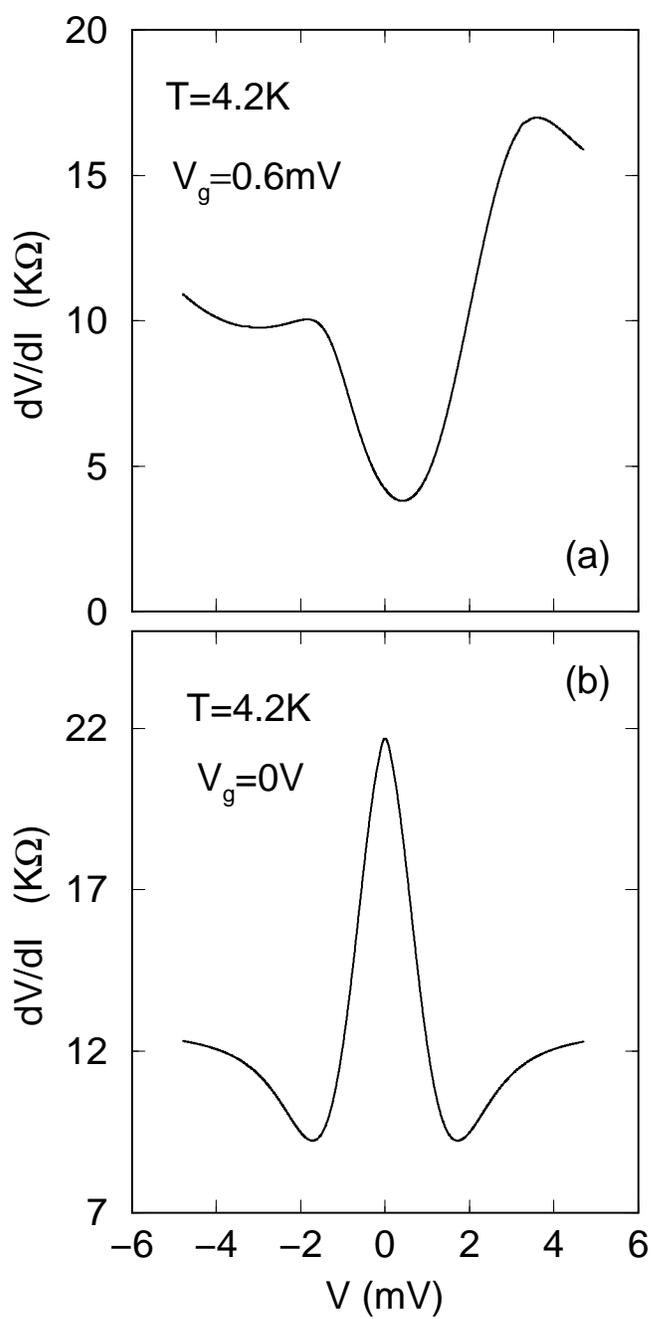
\begin{figure}
\caption{Differential resistance $dV/dI$ as a function of bias voltage
$V$.  (a) For high transparency (on resonance) device corresponding to the
dashed I-V curve of Fig.2.  (b) For low transparency (off resonance) device
corresponding to the solid I-V curve of Fig.2. (5,5) SWNT with length 
$L=19$ unit cells are used.
}
\end{figure}

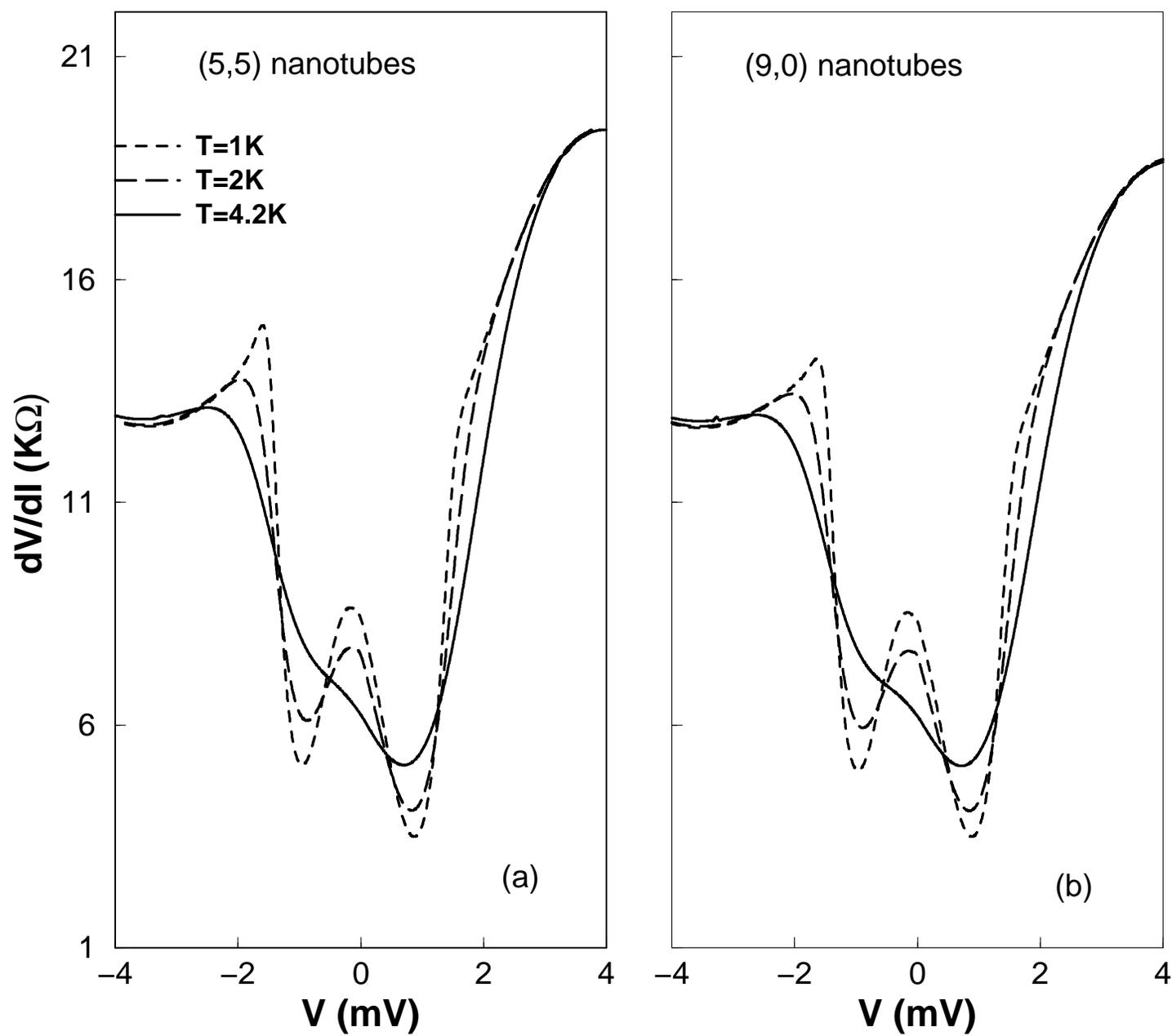
\begin{figure}
\caption{
Differential resistance $dV/dI$ for high transparency devices at different
temperatures. A peak emerges from the overall dip as temperature is lowered.
(a). For (5,5) nanotube devices with $L=19$ unit cells.  (b). For (9,0)
nanotubes with $L=49$.  Other parameters: $V_g=0.6$mV, 
$\Gamma_L=3\times 10^{-3}$, $\Gamma_R=8\times 10^{-3}$.  
}
\end{figure}


\begin{thebibliography}{99}


\bibitem{tsukagoshi}
Kazuhito Tsukagoshi, Bruce W. Alphenaar and Hiroko Ago,
Nature, {\bf 401}, 572(1999).

\bibitem{zhang}
Y. Zhang, {\it et.al.}, Science, {\bf 285}, 1719 (1999).

\bibitem{yu}
A. Yu. Kasumov {\it et.al.}, Science {\bf 284}, 1508(1999).

\bibitem{dai}
A.F. Morpurgo, J. Kong, C.M. Marcus and H. Dai, Science {\bf 286}, 
263(1999).

\bibitem{andreev}
A.F. Andreev, Sov. Phys. JETP {\bf 19}, 1228 (1964).

\bibitem{beena1}
For reviews, See, for example, the article and references therein 
by C.W.J. Bennakker, in {\it Mesoscopic Quantum Physics}, Les Houches, 
Session LXI, 1994.  Eds.  E. Akkermans, G. Montambaux, J.L. 
Pichard and J. Zinn-Justin, (Elsevier Science B.V., 1995);  
Rev. Mod. Phys. {\bf 69}, 731 (1997).

\bibitem{lambert1}
C.J. Lambert and R. Raimondi, J. Phys. Condens. Matter, {\bf 10}, 901
(1998).

\bibitem{beena}
C.W.J. Beenakker, Phys. Rev. B {\bf 46}, 12841(1992);
N.R. Claughton, M. Leadbeater and C.J. Lambert, J. Phys. Condens. Matter
{\bf 7}, 8757 (1995).

\bibitem{ando}
S. Ishizaka, J. Sone and T. Ando, Phys. Rev. B {\bf 52}, 8358 (1995).

\bibitem{yeyati1}
A.L. Yeyati, A. Martin-Rodero and J.C. Curvas, J. Phys. Condens. 
Matter {\bf 8}, 449(1996); Phys. Rev. B {\bf 54}, 7366 (1996);
A.L. Yeyati, {\it et.al.},
Phys. Rev. B {\bf 55}, R6137 (1997).

\bibitem{sun}
Qing-feng Sun, Jian Wang and Tsung-han Lin, Phys. Rev. B. {\bf 59}, 
3831(1999).

\bibitem{wendin}
E.N. Bratus', V.S. Shumeiko and G. Wendin, Phys. Rev. Lett. {\bf 74}, 2110
(1995); Low Temp. Phys. {\bf 23}, 249 (1997); E.V. Bezuglyi {\it et.al.},
Phys. Rev. Lett. {\bf 83}, 2050 (1999).

\bibitem{ruitenbeek}
N. van der Post {\it et.al.},
Phys.  Rev. Lett. {\bf 73}, 2611 (1994); B. Ludoph {\it et.al.}, 
Phys. Rev. B {\bf 61}, 8561 (2000).

\bibitem{kleinsasser}
A.W. Kleinsasser, {\it et.al.}, Phys. Rev. Lett. {\bf 72}, 1738(1994).

\bibitem{kastalsky}
A. Kastalsky, {\it et.al}, Phys. Rev. Lett. {\bf 67}, 3206 (1991); S.J. 
M. Bakker, {\it et.al.}, Phys. Rev. B {\bf 48}, 4168 (1993).

\bibitem{scheer}
E. Scheer {\it et.al.}, Phys. Rev. Lett. {\bf 78}, 3535 (1997).


\bibitem{chico}
L. Chico, {\it et.al.}, Phys. Rev. Lett.
{\bf 76}, 971 (1996); 
V.H. Crespi, M.L. Cohen, and A. Rubio, {\it ibid.} {\bf 79}, 2093 (1997);

\bibitem{jauho}  
A.P. Jauho, N.S. Wingreen, and Y. Meir, Phys. Rev. B 50, 5528(1994).

\bibitem{hatem}
H. Mehrez {\it et.al.}, 
Phys. Rev. Lett. {\bf 84}, 2682 (2000).

\bibitem{bjk}
G.E. Blonder, M. Tinkham and T.M. Klapwijk, Phys. Rev. B {\bf 25}, 4515 
(1982).

\bibitem{marco}
M. Buongiorno Nardelli, Phys. Rev. B. {\bf 60}, 7828 (1999);
A. Rochefort {\it et.al.}, 
Phys. Rev. B {\bf 60}, 13824 (1999).

\bibitem{wan1}
See, for example {\it ab initio} calculations for normal metal leads:
C.C. Wan, {\it et.al.}, Appl. Phys. Lett. {\bf 71}, 419 (1997); 
Jian Wang, {\it et.al.},  Phys. Rev. Lett. {\bf 80}, 4277 (1998).

\bibitem{foot1}
For armchair nanotubes $L$ is measured in terms of unit cells: a unit cell 
is the repeat unit along the armchair tube consisting of two carbon rings.  
For zigzag nanotubes $L$ is in terms of the number of carbon rings. 

\bibitem{rochefort1}
A. Rochefort, D.R. Salahub and P. Avouris, J. Phys. Chem. B {\bf 103}, 641
(1999).

\bibitem{rubio}
A. Rubio, {\it et.al.}, Phys. Rev. Lett. {\bf 82}, 3520 (1999).

\bibitem{venema}
L.C. Venema, {\it et.al.}, Science {\bf 283}, 52 (1999).

\bibitem{luttinger}
Marc Bockrath {\it et al.}, Nature {\bf 397}, 598 (1999); Zhen Yao
{\it et al.}, {\it ibid} {\bf 402}, 273 (1999); Reihold Egger and
Alexander O. Gogolin, Phys. Rev. Lett. {\bf 79}, 5082 (1997); Charles Kane
{\it et al.}, {\it ibid} {\bf 79}, 5086, (1997).

\bibitem{bockrath}
M. Bockrath, {\it et.al.}
Science 275, 1922 (1997).

\end{thebibliography}
\end{document}